\documentclass[11pt]{article}
\usepackage{moriond,epsfig}

\def\be{\begin{equation}}
\def\ee{\end{equation}}
\def\bea{\begin{eqnarray}}
\def\eea{\end{eqnarray}}

\def\cO#1{{\cal O}\left( {#1} \right)}

\def\Tr{{\rm Tr}}
\def\N{{\cal N}}

\begin{document}
\vspace*{4cm}
\title{SEMI-NUMERICAL EVALUATION OF ONE-LOOP CORRECTIONS}

\author{G.~Zanderighi }

\address{Physics Department, Theory Division, CERN, 1211 Geneve 23,
  Switzerland}

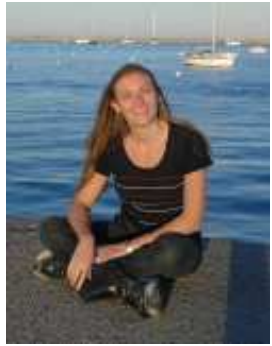
\maketitle\abstracts{ We present a semi-numerical method to compute
  one-loop corrections to multi-leg processes. We apply the method to
  the study of Higgs plus four parton and six gluon amplitudes.}

\section{Motivation}
At the moment, there is no striking discrepancy between Standard Model
(SM) predictions and data. The only missing cornerstone of the SM is
the Higgs boson. However, many hints indicate that the SM is only a
low-energy effective theory.  In view of this, the main tasks of the
LHC will be to discover the Higgs and measure its properties, to
(stress-)test the quantum structure of the SM at the TeV scale and to
unravel possible physics beyond the SM.
At the LHC dominant corrections will be due to higher orders in QCD.
Our ability to accomplish the aforementioned tasks will thus be
limited by the quality of our understanding of QCD.  Next-to-leading
order (NLO) predictions are in this respect important as they allow
one to establish cross-section normalization and shapes and to reduce
the dependence on unphysical factorization and renormalization scales.
In searches for new physics it is important to have a good
understanding of backgrounds, especially for small signal to
background ratios. Finally, loop-induced corrections allow one to get
indirect informations about sectors not directly accessible by
experiments. For instance electroweak precision data lead to an
indirect bound on the Higgs mass.

Despite the fact that at the LHC and ILC most processes/backgrounds
involve multi-particle final states, today very few NLO calculations
of multi-leg processes exist. A remarkable exception is the NLO
calculation of $e^+e^-\to\,$4 fermions~\cite{Denner:2005es}.

A full N-particle NLO calculation requires the tree level prediction
of (N+1) parton amplitudes, the pure virtual correction to the N
parton process, and, since both of the above are divergent, one needs
to compute a set of subtraction terms as well.
While the evaluation of leading-order (LO) amplitudes has been largely
automated and also the computation of one-loop subtraction terms is
well understood, as it requires only the knowledge of the divergent
part, the complexity of the {\em analytical} evaluation of the virtual
terms of multi-leg amplitudes is the limiting factor in NLO
calculations. We therefore developed a method to compute such virtual
corrections semi-numerically.

\section{The method} 
\label{sec:method}
Our semi-numerical method is very simple and general. It is based on
the following three steps:
\begin{itemize}
\item[A:] we one use a combination of Qgraf/Form/Mathematica to
  generate the amplitude for the specific process and to write it as a
  sum of contraction between tensor integrals and coefficients which
  depend on the kinematic of the process (momenta $p_i$ and
  polarizations $\varepsilon_i$)
\begin{equation}
{\cal A}(p_1,\dots p_N) = \sum K_{\mu_1\cdots\mu_M}
  (p_1,\ldots p_N; \varepsilon_1,\ldots \varepsilon_N) I^{\mu_1\dots
    \mu_M}(D; \nu_1,\dots \nu_N), 
\end{equation}
where $I^{\mu_1 \cdots \mu_M} (D; \nu_1,\ldots,\nu_N)$ denotes the
generic scalar integral~\footnote{Note that we consider here only the
  case of massless internal propagators although the method is more
  general.}
\begin{equation}
I^{\mu_1 \cdots \mu_M} (D; \nu_1,\ldots,\nu_N) \equiv \int
\frac{d\,^D l}{i \pi^{D/2}}\ \frac{l^{\mu_1}\cdots l^{\mu_M}}
{d_1^{\nu_1} d_2^{\nu_2} \cdots d_N^{\nu_N}}, 
\quad 
d_i \equiv (l+q_i)^2\,,\quad
q_i \equiv \sum_{j=1}^i p_j\,. 
\end{equation}
Apart from trival manipulations, the critical point here is to cancel
all quadratic terms in the loop-momentum in $I^{\mu_1 \dots \mu_M}$,
so that $K_{\mu_1 \dots \mu_M}$ does not depend on the $D$-dimensional
metric tensor;
\item[N1:] using the Davydychev reduction~\cite{Davydychev:1991va}, we
  reduce tensor integrals to a combination of higher dimensional
  scalar integrals with shifted exponents;
\item[N2:] using a complete set of recurrence
  relations~\cite{Giele:2004iy,Ellis:2005zh}, we reduce each scalar
  integral to a combination of analytically known, basis integrals. A
  sample relation reads
\begin{equation}
\label{eq:samplerel}
I(D;\{\nu_l\})=
\frac{1}{{B}(D-1-\sigma)}
\left(I(D-2;\{\nu_l\}) -\sum_{i=1}^Nb_iI(D-2;\{\nu_l-\delta_{li}\})
\right)\,,
\end{equation}
where $S_{ij}=\left(q_i-q_j\right)^2$, $b_i\equiv\sum_{j=1}^N
S_{ij}^{-1}$ and $B\equiv\sum_{j=1}^N b_i$.
\end{itemize}
While step A is done analytically only once for each process, steps N1
and N2 are repeated numerically for each phase space point. Note that
since recursion relations generally involve $D= 4-2\epsilon$, results
for intermediate scalar integrals are stored as Laurent expansions in
$\epsilon$. Since one uses analytical expressions for basis integrals,
unless there is a specific numerical instability, no loss of accuracy
in the semi-numerical evaluation is to be expected.  However, it is
well known that for so-called exceptional momentum configurations
normal recursion relations become unstable, e.\ g.\ 
eq.~(\ref{eq:samplerel}) is undefined for $B=0$ (other ones for
$\det(S)=0$).  These phase space points correspond to either
accidental degeneracies (such as planar configurations) or to physical
thresholds. They have vanishing phase space measure but recursions
become numerically unstable close to those points.
The solution we adopt here is an extension~\cite{Ellis:2005zh} of the
method suggested in~\cite{Giele:2004ub} and tested there with the
simple forward $\gamma \gamma \to \gamma \gamma$ scattering. The idea
is to exploit the existence of a small parameter ($B$ or $\det (S)$)
to define expanded relations.  For instance, for small $B$
eq.~(\ref{eq:samplerel}) can be expanded in $B$ to give 
\begin{equation}
I(D-2;\{\nu_l\}) = \sum_{i=1}^N b_i I(D-2;\{\nu_l-\delta_{li}\})
+ \left(D-1-\sigma\right)\,B\,I(D;\{\nu_l\})\,. 
\end{equation}
The structure of the expanded relations must be such that the integral
to the l.\ h.\ s.\ side is expressed as a sum of integrals with
coefficients $\cO{1}$, which have a simpler kinematical matrix and
will generally not give rise to further exceptional expansions (though
in some cases they might) and a sum of integrals whose coefficients
are suppressed by a small parameter but who have exactly the same
kinematical matrix and will therefore be computed by applying the same
expanded relation iteratively.  Similarly, if $\det(S)\ll1$ one
expands around the eigenvector corresponding to the smallest
eigenvalue.
Since exceptional points are easily detected numerically, there is no
need to understand analytically their origin.
This is particularly important if one considers cases with internal
masses in the loop.
In this respect, the method~\cite{Denner:2005nn} used to compute treat
exceptional regions for instance in the calculation of $e^+e^-\to$ 4
fermions is very similar to ours, but relies on a different tensor
reduction.

\section{Applications}
\subsection{Higgs plus four partons} 
As a first application of the method we considered the NLO corrections
to Higgs plus four parton amplitudes in the large $m_t$ limit. Gluon
fusion Higgs plus dijet production is to be considered a background to
vector boson fusion Higgs plus dijet process. Indeed the latter is the
most promising channel to measure the Higgs couplings at the LHC,
while the former is plagued by large QCD uncertainties. It was
therefore important to compute this process at NLO in QCD. We first
computed the virtual corrections numerically~\cite{Ellis:2005qe}.
As a check of the calculation, we computed also the analytical
amplitude for amplitudes with four external quarks. In the two-quark
two-gluon and four-gluon amplitudes we could check the poles, the Ward
identities, as well as relations involving various color amplitudes
(cyclicity, reflection and decoupling identities~\cite{Bern:1990ux}).
For all non-exceptional points examined~\footnote{After rescaling the
  hard event to the hard scale of the process, we choose to deem an
  event to be exceptional if $B$ or $\det(S)$ is less then
  $\varepsilon$ and played with values of $\varepsilon$ in the range
  $10^{-3}-10^{-6}$.}  the relative accuracy turns out to be of
$\cO{10^{-13}}$.
We also examined the stability close to those exceptional points. Here
one defines a target accuracy $\rho$ (we choose $\rho = 10^{-6}$) and
stops the iterative expansion once this accuracy reached. Our results
turned out to be always within the target accuracy.
To complete the calculation of cross section for this process one
needs to implement real radiation and compute subtraction terms.
Preliminar results have been presented in~\cite{KELL}.

\subsection{Six gluon amplitudes} 
As a next application we considered six gluon amplitudes.  One
important difference is that starting with six parton process, after
eliminating one momenutm due to momentum conservation the remaining
partons are no longer independent.
One expects therefore the Davydychev decomposition to becomes rapidly
inefficient with increasing number of legs, since one decomposes in a
redundant set of momenta.  We also found a quite large loss in
numerical precision using Davydychev reduction starting from rank four
six-point tensor integrals.  Therefore we choose to use a method which
explicitly uses the (over-)completeness of the set of
momenta~\cite{egz3}.
We performed a standard color decomposition of the amplitude ($h_i$
and $a_i$ denote helicity and color of gluon i)~\cite{Bern:1990ux}
\begin{eqnarray}
{\cal A}_n ( \{p_i,h_i,a_i\} ) & = & g^n
\left( 
  \sum_{\sigma \in Z_n} 
N_c \Tr( T^{\sigma(a_1)}\cdots T^{\sigma(a_{n})} ) \,A_{n;1} \right. \\
&+& \left. 
      \sum_{c=2}^{\lfloor{n/2}\rfloor+1}
      \sum_{\sigma \in S_n/S_{n;c}}
     \Tr( T^{\sigma(a_1)}\cdots T^{\sigma(a_{c-1})} )\, 
     \Tr ( T^{\sigma(a_c)}\cdots T^{\sigma(a_n}) \,A_{n;c}\right)\,. 
\end{eqnarray}
Subleading color amplitudes $A_{n;c}$ ($c >1$) are completely
determined by the leading color ones $A_{n;1}$, therefore we
considered only $A_{n;1}$ in the basic ordering of momenta.
One can then consider a helicity decomposition of the amplitude. Out
of 64 possible amplitudes only 8 are independent, the others can be
obtained by parity and cyclicity.  Specifically we choose to consider
the two finite amplitude $\{6+\}$ and $\{1-5+\}$, the three MHV
amplitudes $\{2-4+\}$ and the three NMHV amplitudes $\{3-3+\}$.
In order to compare with the literature we considered a supersymmetric
decomposition of the amplitude, i.\ e.\ apart from the spin 1
($A^{[1]}$) and spin 1/2 ($A^{[1/1]}$) terms, we considered also the
contribution of a complex scalar in the loop ($A^{[0]}$) and used this
to construct amplitudes with ${\cal N}=4$ and ${\cal N}=1$
supersymmetric multiplets in the loop
\begin{eqnarray}
{A}^{{\N}=4}&=&A^{[1]}+4 A^{[1/2]}+3 A^{[0]}\,, \nonumber \\
{A}^{{\N}=1}&=& A^{[1/2]}+A^{[0]}\,. 
\end{eqnarray}
After the revision of some analytical calculations we now agree with
all public available results (for details about the comparison with
analytical results see~\cite{egz3}). The relative accuracy is now of
the order of $\cO{10^{-8}- 10^{-9}}$. The main new results are 5 out
of 8 scalar amplitudes, which were still unknown analytically. This
completes the virtual calculation of six-gluon amplitudes.\footnote{We
  note that contemporary to our paper in~\cite{Britto:2006sj} all
  cut-constructable parts to six gluon amplitudes have been computed
  and that very recently one more NMHV scalar six-gluon amplitude has
  been computed~\cite{Berger:2006ci}.}
This calculation illustrates the complementarity between analytical
and numerical results: the scalar contribution is numerically the
easiest one, while it is by far the hardest contribution to compute
analytically.  Additionally, these results show that numerical methods
provide very useful independent checks of analytical calculations.

Despite the fact that virtual corrections to six gluon amplitudes are
now fully know numerically, the way to go to obtain cross sections is
still long: one should extend the expanded relations to cases with six
external particles similarly to what has been done for five-point
amplitudes (alternatively, one could use a more brute-force approach
of interpolating around the exceptional points), one should consider
amplitudes with external quarks, real radiation from a six parton
ensemble and compute subtraction terms. Finally, one should merge all
these elements in an efficient phase space integrator.

\section*{Acknowledgments}
I thank Keith Ellis and Walter Giele for collaboration on this
project. 


\section*{References}
\begin{thebibliography}{99}

\bibitem{Denner:2005es}
  A.~Denner, S.~Dittmaier, M.~Roth and L.~H.~Wieders,
  Phys.\ Lett.\ B {\bf 612} (2005) 223 and Nucl.\ Phys.\ B {\bf 724} (2005) 247. 

\bibitem{Davydychev:1991va}
  A.~I.~Davydychev,
  Phys.\ Lett.\ B {\bf 263}, 107 (1991).

\bibitem{Giele:2004iy}
  W.~T.~Giele and E.~W.~N.~Glover,
  JHEP {\bf 0404}, 029 (2004).

\bibitem{Ellis:2005zh}
  R.~K.~Ellis, W.~T.~Giele and G.~Zanderighi,
  Phys.\ Rev.\ D {\bf 73}, 014027 (2006). 

\bibitem{Giele:2004ub}
  W.~Giele, E.~W.~N.~Glover and G.~Zanderighi,
  Nucl.\ Phys.\ Proc.\ Suppl.\  {\bf 135}, 275 (2004).

\bibitem{Denner:2005nn}
  A.~Denner and S.~Dittmaier,
  Nucl.\ Phys.\ B {\bf 734}, 62 (2006).

\bibitem{Ellis:2005qe}
  R.~K.~Ellis, W.~T.~Giele and G.~Zanderighi,
  Phys.\ Rev.\ D {\bf 72}, 054018 (2005). 

\bibitem{KELL}
See talk of K. Ellis at Loops and Legs 2006. 

\bibitem{egz3}
  R.~K.~Ellis, W.~T.~Giele and G.~Zanderighi,
  arXiv:hep-ph/0602185.


\bibitem{Bern:1990ux}
  Z.~Bern and D.~A.~Kosower,
  Nucl.\ Phys.\ B {\bf 362} (1991) 389.

\bibitem{Britto:2006sj}
  R.~Britto, B.~Feng and P.~Mastrolia,
  Phys.\ Rev.\ D {\bf 73} (2006) 105004. 


\bibitem{Berger:2006ci}
  C.~F.~Berger, Z.~Bern, L.~J.~Dixon, D.~Forde and D.~A.~Kosower,
  hep-ph/0604195.


\end{thebibliography}
\end{document}